\documentclass[aps,prl,twocolumn,superscriptaddress,10pt]{revtex4-2}
\usepackage[dvips]{graphicx}
\usepackage{epsfig}
\usepackage{natbib}
\usepackage{amssymb}
\usepackage{url}
\usepackage{pslatex}
\usepackage{color}
\usepackage{times,fancyhdr}
\usepackage{amsmath}
\usepackage{amsfonts}
\usepackage{amsthm,amscd}
\usepackage{amsbsy}
\usepackage{latexsym}
\usepackage{bm}
\usepackage{layout}
\usepackage{makeidx}
\usepackage{epstopdf}	
\usepackage{color}
\usepackage{hyperref}
\usepackage[latin1]{inputenc}
\usepackage[T1]{fontenc}
\usepackage[ddmmyyyy]{datetime}
\usepackage{courier}	
\def\be{\begin{equation}}
\def\ee{\end{equation}}

\def\be{\begin{equation}}
\def\ee{\end{equation}}

\def\({\left(}
\def\){\right)}

\newdateformat{mydate}{\THEYEAR-\twodigit{\THEMONTH}-\twodigit{\THEDAY}}
\begin{document}
\title{Thermal radiative cooling of carbon cluster cations C$_N^+$, 
$N = 9, 11,12, 17-27$}
\author{Shimpei Iida}
\affiliation{Tokyo Metropolitan University, Tokyo 192-0397, Japan}
\author{Wei Hu}
\affiliation{School of Science, Tianjin University, 92 Weijin Road, Tianjin 300072, China}
\author{Rui Zhang}
\affiliation{School of Science, Tianjin University, 92 Weijin Road, Tianjin 300072, China}
\author{Piero Ferrari}
\affiliation{Quantum Solid-State Physics, Department of Physics and Astronomy,
KU Leuven, 3001 Leuven, Belgium}
\author{Kei Masuhara}
\affiliation{Tokyo Metropolitan UniversitY, Tokyo 192-0397, Japan}
\author{Hajime Tanuma}
\affiliation{Tokyo Metropolitan UniversitY, Tokyo 192-0397, Japan}
\author{Haruo Shiromaru}
\affiliation{Tokyo Metropolitan UniversitY, Tokyo 192-0397, Japan}
\author{Toshiyuki Azuma}
\affiliation{Atomic, Molecular and Optical Physics Laboratory, RIKEN, Saitama 351-0198, Japan}
\author{Klavs Hansen}
\email{klavshansen@tju.edu.cn,hansen@lzu.edu.cn}
\homepage{http://www.klavshansen.cn/}
\affiliation{Center for Joint Quantum Studies and Department of Physics, 
School of Science, Tianjin University, 92 Weijin Road, Tianjin 300072, China}
\affiliation{Lanzhou Center for Theoretical Physics, Key Laboratory of Theoretical Physics of Gansu Province, 
Lanzhou University, Lanzhou, Gansu 730000, China}

\mydate
\date{\today,~\currenttime}

\begin{abstract}
The radiative cooling rates of C$_N^+$ clusters 
($N = 9, 11, 12, 17-27$) have been measured in the ultrahigh vacuum 
of an electrostatic storage ring to values on the order of $10^4$ s$^{-1}$.
The rates were measured as a competing channel to unimolecular decay, 
and the rate constants pertain to the excitation energies where these two channels 
compete.
Such high values can only be explained as photon emission from thermally 
excited electronic states, a mechanism that has also been seen in 
polycyclic aromatic hydrocarbon cations.
The high rates have a very strong stabilizing effect on the clusters and the 
underlying mechanism gives a high energy conversion efficiency, with the 
potential to reach high quantum efficiencies in the emission process. 
The competing decay of unimolecular fragmentation defines upper 
limits for photon energies that can be down-converted to lower 
energy photons.
Including previously measured cluster sizes provides the limits for all clusters 
C$_N^+$, $N=8-27$, of values that vary from 10 to 14.5 eV, with a 
general increase with size.
Clusters absorbing photons of energies below these limits cool down 
efficiently by emission of photons via electronic transitions and their 
fragmentation is strongly reduced, increasing their survival in HI regions.
\end{abstract}
\maketitle

\section*{Introduction}

With the large and growing number of molecular species 
identified in interstellar space (\cite{TielensRMP2013, McGuireAJSS2018}), 
chemistry in vacuum is becoming increasingly important.
A key question concerns the formation and survival of these molecules as well 
as the related question of the propensity for emission of photons from highly 
excited species.

The mechanisms by which internal excitation energy is dissipated play a 
crucial role for the survival rate of highly excited molecules and clusters and 
thereby also in the quantitative description of the abundances of carbon-containing 
compounds, such as clusters and in particular polycyclic aromatic hydrocarbon 
(PAH) molecules (\cite{MontillaudAA2013,BerneAA2015,AndrewsAA2016}).
In the virtually collision-free interstellar environment, dissipation of 
molecular excitation energy occurs overwhelmingly by unimolecular reactions or 
by radiative cooling, which makes studies of decays in ultrahigh vacuum 
storage rings and similar devices ideal for understanding such processes.

The radiative cooling from molecules in the interstellar medium has 
been discussed extensively in terms of various emission processes occurring
 in the infrared (e.g. \cite{HollenbachRMP1999}).
An important alternative channel which is attracting increasing 
attention is photon emission from thermally populated electronic states.

For historical reasons, this mechanism is denoted recurrent fluorescence (RF).
Also the term Poincar{\'e} radiation is used. 
RF does not require that the initial injection of excitation energy is a result of 
photon-absorption, although this excitation mode is expected to be common 
under interstellar conditions. 
The energy released in the formation of molecules in collisional attachments 
may equally well provide the necessary excitation energy.
The effect was suggested several decades ago (\cite{Nitzan1979}) and in 
astrophysical context in \cite{Leach1987,Leger1988}.
A number of laboratory experiments have already been interpreted in terms of 
this phenomenon, both for carbon containing species and for several metal 
and semiconductor clusters; see Refs. \cite{JUAprl1996,AndersenEPJD2001,
Chandrasekaran2014,Kono2015,ItoPRL2014,KonoPRA2018} for experiments
on carbon anions.
Of particular astrophysical interest are the applications where RF was invoked 
to explain the quenching of the decays of PAH molecules 
(\cite{martin13,BernardNIMB2017,MartinPRA2019,StockettJCP2020}).

Studies of radiative cooling of several of the cationic carbon 
clusters in the homologous series reported here were given in 
\cite{ChenPCCP2019}.
Also the direct detection of photons emitted from size selected clusters 
in such processes has been accomplished with the observation of thermal, visible 
photons emitted from C$_6^-$ and C$_4^-$ 
(\cite{EbaraPRL2016,YoshidaJPCS2017}) and of  naphthalene cations 
(\cite{SaitoPRA2020}).
Ref. \cite{FerrariIRPC2019} provides a review of the status of the laboratory 
studies of the subject until 2019.
The recent report in \cite{RademacherJPCA2022} on the absorption spectra of 
carbon clusters corroborates the interpretation of the quenching as due to RF.
The work reports absorption peaks of even-numbered cationic carbon clusters 
of sizes $n=12-26, 28$ in the range from a little below 1 eV for the largest to 
a little above 2 eV for the smallest. 

One astrophysical aspect of interest of these studies is the possibility of a 
connection to the extended red emission 
(\cite{CohenAJ1975,SchmidtAJL1980,FurtonAPJ1990,FurtonAPJ1992,WittASS2020}).
Another is the still largely open question of the origin of the diffuse 
interstellar bands and their carriers (\cite{SalamaApJ2011,OmontAA2016,
APJonesRSOS2016}),  and the possible link to the carriers of the extended red 
emission (\cite{LaiMNRAS2020}). 

The emission of RF photons occurs after excitation to quantum states 
at energies that usually exceed thermal energies by a large factor,
with a concomitant reduction in the populations of such excited states.
The much higher oscillator strength of electronically excited states can
nevertheless favor emission from these states over the slower photon 
emission via vibrational transitions. 
As a result, RF emission rates can reach values more than a 
factor thousand above those of infrared (IR) emission, making RF a 
highly competitive stabilizing channel for those molecules that exhibit the 
phenomenon, as well as an important source of photons in the near-IR and 
visible parts of the spectrum.
The shortest measured time constants are those of metal clusters, 
but also carbon clusters can show very high photon emission rates, as 
demonstrated by the ten $\mu$s time constant observed for C$_{4}^-$ 
(\cite{Kono2015}) and the slightly longer values for cationic fullerenes 
(\cite{TomitaPRL2001}).

The time constants of RF emission obviously depend strongly on the 
specific electronic structure of the radiating molecules via the energy of 
the excited and emitting state, and the observed values not 
surprisingly also vary widely.
The most striking example of this chemical dependence is the radiative 
constants of small carbon anions.
Both C$_5^-$ and C$_7^-$ radiate with time constants in the tens of ms range 
in vibrational transitions
(\cite{GotoJCP2013,NajafianJCP2014}), whereas both C$_4^-$ 
and C$_6^-$ have time constants around ten $\mu$s (\cite{Kono2015,ItoPRL2014}).
Equally striking is the effect of addition of a single hydrogen atom to 
C$_6^-$ to form C$_6$H$^-$.
This reduces the photon emission rate to values similar to those of 
C$_5^-$ and C$_7^-$ (\cite{ItoPRL2014}) and pentacene 
(\cite{IidaPRA2021}).

The measurements reported here were performed with an electrostatic 
storage ring.
They made use of a special feature of most cluster sources that produce 
clusters or molecules in highly excited states, viz. that the energy distributions 
have widths that give rise to a wide continuum of rate constants.
This continuum must then be considered in the description of the decay rates.
In spite of the underlying statistical and completely exponential 
unimolecular decay rate constants, the measured decay rates, which are 
ensemble averages over these broad energy distributions, will for this 
reason give rise to non-exponential decay rates.
Under reasonable general conditions these rates are well described by a 
power law, in the simplest form as $1/t$.
The effect is well understood theoretically and has been discussed at 
length in the literature (see for example \cite{Andersen2002}) and 
documented experimentally (see \cite{HansenPRL2001} for the first 
experimental observation).

The simple power law requires that the experimentally detected 
unimolecular decay is the dominant energy loss channel.
Photon emission will introduce an exponential suppression of the $1/t$ 
decay profile.
The exponential suppression arises due to the different energy dependence 
of the two decays.
A brief derivation is given below, and Ref. \cite{FerrariIRPC2019} 
can also be consulted for an analysis of this question.
Specific examples of calculated rate constants for C$_4^-$ and C$_6^-$, 
based on known spectroscopic and thermal properties of the clusters are 
given in \cite{Kono2015,ChandraJCP2017}.

The exponential suppression of the power law decay can be used to 
determine quantitatively the radiative time constant (\cite{HansenJCP1996,JUAprl1996,martin13}).
The radiative time constant measured is the one pertaining to the energy 
where the two curves for the unimolecular decay and the photon emission 
cross (see Fig. 3 in \cite{Kono2015} for an example).
The measurements will give the quenching rate constant at this 
crossing energy with little need for any further characterization of the two 
decay rate constants involved.
A majority of the determinations of the radiative rate constants in the 
references mentioned above has applied this technique.
The determination of the energy content of the crossing point, however,
requires an expression for the energy dependence of the unimolecular rate 
constant.

The experiments reported here measured the radiative quenching 
time for a number of cationic carbon clusters of medium size.
They extend the measurements of the limited size range reported 
in \cite{ChenPCCP2019} to a number of cluster sizes not previously 
produced in sufficient intensities to make the experiments feasible.
For clusters C$_{11}^+$ and C$_{19}^+$ the measurements of 
spontaneous 
unimolecular decay quenching was supplemented by measurements 
of the decay rates induced by one-photon absorption at varying 
storage times, providing a test of the assumptions of a statistical 
decay and broad initial cluster energy distributions.

\section{Experimental procedure}

The clusters were produced in a laser ablation source by a 532 nm 
light pulse from a Nd:YAG laser hitting a rotating surface of graphite powder.
The graphite feedstock was isotopically purified to $^{12}$C, and all 
C$_N^+$ were mass-selected with sufficient resolution
to separate $^{12}$C$_{N}^{+}$ from $^{12}$C$_{N}$H$^{+}$.
No cooling gas was applied, and the ions were consequently produced highly 
excited with broad internal energy distributions.
The source has previously been used for the production of a number of molecules 
and clusters with consistently reliably results for the width of the energy 
distribution, both for molecules desorbed intact and for clusters created during the 
ablation, as the clusters in the present study, and with both negative 
and positive charge.
The relevant figure of merit in this context is the constancy and reproducibility of 
the time profiles of the spontaneous decays of the stored ions.

\begin{figure}
\begin{center}
\vspace{-1cm}
\includegraphics[width=0.7\columnwidth,angle=90]{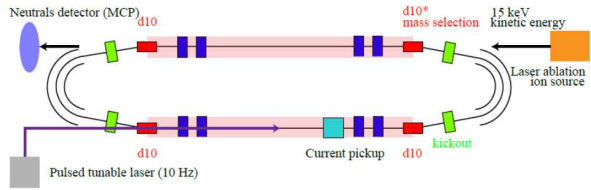}
\vspace{-1cm}
\caption{\label{TMUEring}
A sketch of the electrostatic storage ring used in these studies,
the TMU E-ring.
The ions were produced in the laser ablation source shown 
in the upper right corner of the figure, accelerated to 15 keV and 
injected into the ring.
The mass selection was accomplished by pulsing the potential on the 
10 degree deflection plate d10* and on the kickout electrode.
The neutral fragments that were produced during storage by decays 
in the top straight section were detected with the neutral particle 
detector at the end of that section on the top left of the figure.
For the laser excitation experiments, a light pulse from a tunable 
Optical Parametric Oscillator (OPO) laser was introduced at the lower 
left hand corner with the light propagating toward the lower right 
corner along the straight section of the ring.
After measurement, the ions were ejected by switching off the d10* 
deflection plates, which also prepared for the next injection.}
\end{center}
\end{figure}

After production in the source, the ions were accelerated vertically 
to 15 keV, turned to horizontal motion and injected into the 7.7 m 
circumference storage ring, shown in Fig. \ref{TMUEring}, by switching 
the set of 10 degree deflection plates in the ring closest to the cluster source.  
All cations produced in the source were accelerated and injected into 
the ring.
The cluster size of interest was selected by pulsing the set of 10 
degree deflection plates closest to the injection gate twice.
First time on injection and the second time when the different mass ions were
spatially separated after a number of revolutions in the ring. 
Ref. \cite{Jinno2007} gives more details of this procedure and of the ring.

The decays were measured time-resolved turn-by-turn with a neutral 
particle detector.
The detector hence monitored the quasi-instantaneous decay rate of the 
stored species, and \textit{not} the surviving stored ion populations.
The detector is located at the end of the straight section of the injection side.
An injection cycle ends with the ions being dumped after 90 or 10 ms of 
storage, depending on the decay rate, and a new cycle started.

The signal measured in the neutrals detector comprises all channels that 
emit neutral massive particles, irrespective of whether these are single 
atoms or molecules, and the measurements did not identify the decay product.
The precise unimolecular decay channels are of interest for the energy 
content of the radiating clusters and the information on these channels 
available from the literature is therefore reviewed briefly in the 
discussion section.

The spontaneous decays were measured for cluster sizes $N=9,11,12,17-27$, 
and for laser enhanced decays also for $N=11, 19$ (analogous results for 
$N=8,13-16$ were already reported in \cite{ChenPCCP2019}).
The number of ion injection cycles varied with size and reached 140 000 for 
$N=27$ in order to reduce the statistical noise to a level comparable 
to or below the so-called betatron oscillations, which is the other main source 
of turn-by-turn fluctuations in the spectra. 

\begin{figure}
\begin{center}
\includegraphics[width=0.7\columnwidth,angle=90]{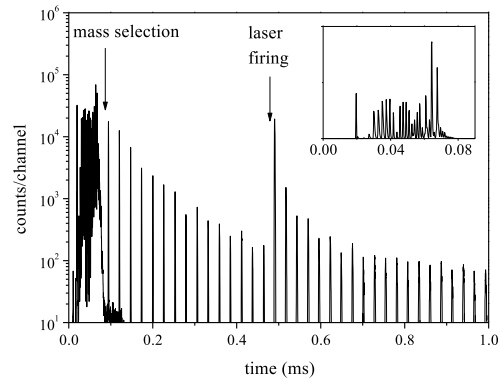}
\caption{\label{laserexample}
The C$_{11}^+$ spontaneous and photo-induced decay rate
after absorption of a 520 nm photon at the time indicated
by the rightmost arrow.
For these experiments a second detector located diagonally across 
from the primary was also used, and the circulation time is 
therefore twice the time separation of peaks seen in this figure.}
\end{center}
\end{figure}

Fig. \ref{laserexample} shows an example of the spontaneous decay rate of 
C$_{11}^+$ for the first few turns in the ring.
The ions were injected in a short bunch and the turn-by-turn decays therefore appear 
as peaks, corresponding to the passage of the bunch through the 
detector side of the ring.
The neutral counts seen in Fig. \ref{laserexample} up to 80 $\mu$s are decays 
of all species produced in the source and transferred to and stored in the ring 
before the final mass selection.
The peaks seen between 25 and 70 $\mu$s are all from clusters separated by 
one carbon atomic mass.
The mass selection pulse was applied around 80 $\mu$s.
Note that the spectrum represents decay rates and cannot be compared 
directly with spectra observed in mass spectrometers. 

The figure also shows the enhanced counts caused by photo-excitation which 
here occurs just before 0.5 ms.
Photo excitation was done with a single nanosecond pulse from a 10 Hz 
tunable optical parametrical oscillator laser. 
The wavelength was fixed at 520 nm (2.38 eV) and for $N=19$ the laser was 
fired at times $9.7 + 10 k$ ms, with $k$ an integer from 0 to 7.
For $N=11$ the laser was fired at times 0.49, 1.02, 1.50, 2.02, 2.50, and 3.03 ms.  
The laser pulse energy was 1.0 mJ/pulse in all cases.
The pulse energy was low enough to cause single photon absorption only.

\section{Results and analysis}

\begin{figure}
\begin{center}
\includegraphics[width=\columnwidth,angle=0]{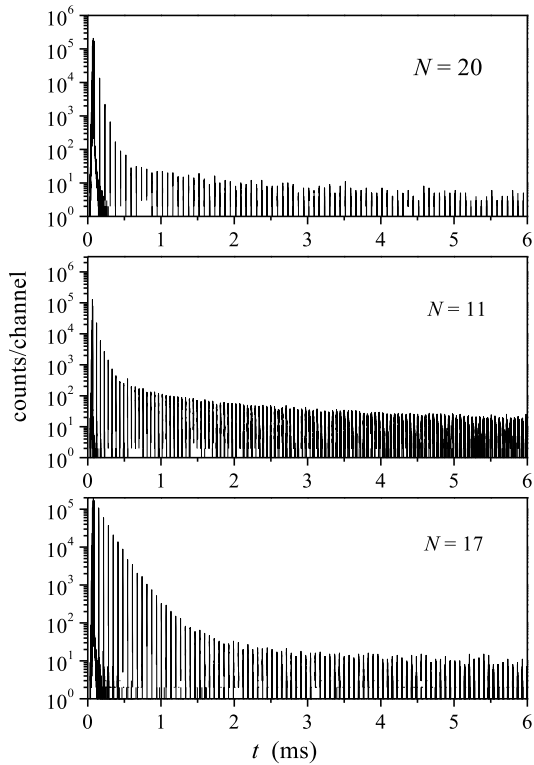}
\vspace{-1cm}
\caption{\label{RawSpec}
The first six milliseconds of the measured spectra of, from top 
to bottom, C$_{20}^+$, C$_{11}^+$, and C$_{17}^+$.
The flat, almost constant parts of the spectra at the long times 
are due to rest gas collisions.  
The different cooling times are clearly seen as different 
cutoff times.}
\end{center}
\end{figure}

Fig. \ref{RawSpec} shows spontaneous decay spectra of the clusters 
C$_{20}^+$, C$_{11}^+$, and C$_{17}^+$ with visibly different radiative 
cooling constants.  
After peak integration, the background induced by the collisions with 
the rest gas was subtracted.
This was done by fitting an exponentially decreasing function to the measured 
signal at the last half of the storage time where the spontaneous decay had 
effectively ceased and only the collision-induced decay was present.
Subtracting this background provided the spontaneous decay rates as a function 
of time. 

In order to relate the observed decay rates to radiative cooling rate constants, 
we reiterate the derivation made in \cite{Kono2015}.
The measured signal, $I_{\rm d}$, is proportional to the integral over the energy 
distribution of the product of the unimolecular rate constant and the population.
For each energy, the latter is exponentially depleted by both the unimolecular 
decay and the photon emission, with the energy dependent rate constants, 
$k_{\rm d}(E)$ and $k_{\rm p}(E)$, respectively.
Hence the measured unimolecular rate must be calculated as the integral (assuming 
a flat initial energy distribution):
\begin{equation}
I_{\rm d} \propto \int_0^{\infty} k_{\rm d}(E) 
{\rm e}^{-\left(k_{\rm d}(E) +k_{\rm p}(E)\right)t}{\rm d} E.
\end{equation}
The energy dependence of $k_{\rm p}$ is weak, in contrast to the very 
strong energy dependence of $k_{\rm d}$. 
This is an effect of the fact that the frequency factor of the photon 
emission rate constant is limited by the Thomas-Reiche-Kuhn (TRK) sum 
rule and in practice is much smaller than that upper limit.
This makes it many orders of magnitude smaller than that of the unimolecular decay.
If photon emission is a competitive channel, this imposes the condition that also the 
ratio of the respective activation energies is significantly smaller than unity, i.e. that
the photon energy is much smaller than the evaporative activation energy.
These aspects are discussed in detail in \cite{FerrariIRPC2019}.
With this general argument, the energy dependence of $k_{\rm p}$ can therefore 
be ignored compared to that of $k_{\rm d}$ and set to a constant in the integral, 
and the exponential containing it extracted from the integral:
\begin{equation}
I_{\rm d} \propto {\rm e}^{-k_{\rm p} t} \int_0^{\infty} 
k_{\rm d}(E) {\rm e}^{-k_{\rm d}(E) t}{\rm d} E.
\end{equation}
The integral is identical to that of the non-radiative decay rate. 
As calculated in several places, see e.g. \cite{HansenPRL2001}, it is proportional to 
$1/t$ and hence
\begin{equation}
\label{Id}
I_{\rm d} \propto \frac{{\rm e}^{-k_{\rm p} t}}{t}.
\end{equation}
For spontaneous decays, the time of ablation in the source is the zero of this time.
In the laser excitation experiments, the absorption of the photon reheated the ions
and thus restarted the power law decay with the laser firing time, 
$t_{\rm las}$, as the redefined time zero.
For small photon energies, the reheating may only be partial, as seen in 
\cite{SundenPRL09}.
However, fits of the photo-enhanced data with the equation 
\begin{equation}
\label{Idlas}
I_{\rm d} \propto \frac{{\rm e}^{-k_{\rm p} t}}{t-t_{\rm las}}
\end{equation}
showed that absorption of the 520 nm photon effectively did reset 
the time to zero in these experiments.

The derivation sketched above requires that the emission of a single photon is 
sufficient to quench the unimolecular decay.
This is the most likely scenario, given the electronic origin of the photons and 
the size of the clusters.
The alternative is that the energy of the emitted photons is too 
low to quench the unimolecular decay.
Then photon emission must instead best be described as continuous cooling.
This will also produce an exponential suppression of 
the decay, albeit with a slightly different time dependence 
(\cite{FerrariIRPC2019}).
In either case, the experimentally measured $k_{\rm p}$ is still the 
relevant quenching rate constant. 

\begin{figure}
\begin{center}
\includegraphics[width=0.8\columnwidth,angle=90]{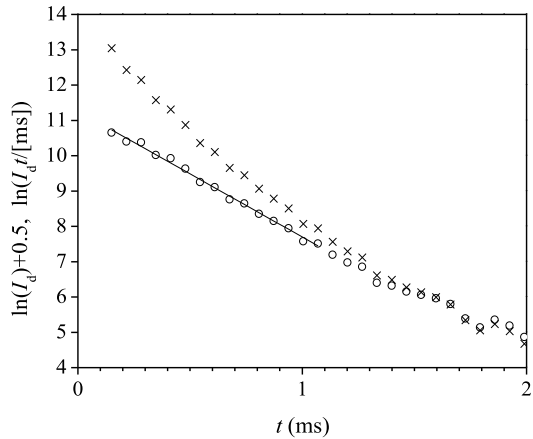}
\vspace{-1cm}
\caption{\label{SLfits} The logarithm of the integrated peak intensity, 
$I_{\rm d}$ (crosses), and of the product $t I_{\rm d}$ (circles) for the 
spontaneous decay of the $N=17$ cluster, chosen for its slow decay.
The curvature on the $I_{\rm d}$ trace is clearly visible, whereas the 
$I_{\rm d} t$ trace shows the expected straight line behavior.
The straight line is the fit of $t I_{\rm d}$.  
The curves reach a noise floor at $10^{-3}$ of the initial signal a little 
above 2 ms.
For display purposes $\ln(I_{\rm d})$ has been shifted up by 0.5.}
\end{center}
\end{figure}

Proceeding with the assumption of large photon energies, the data have 
been fitted with the expression in equation~(\ref{Id}) or equation~(\ref{Idlas}), as 
applicable.
To extract $k_{\rm p}$, it is most convenient to rewrite the expressions 
as $\ln(I_{\rm d} t)$ and $\ln(I_{\rm d} (t-t_{\rm las}))$ and fit a straight 
line.
An example of such a fit is shown in Fig. \ref{SLfits}, together 
with the semilog plot of the unmodified count rate for comparison.
For the laser excitation experiments, the photo enhanced signal 
was the relevant $I_{\rm d}$.
It was extracted from the data by subtracting a reference spectrum recorded 
without photo-excitation and normalized to identical intensity by the 
pre-laser counts.

The validity of the analysis requires that these photo-enhanced signals 
have the same values of the fitted $k_{\rm p}$, independently of the 
storage time before photon absorption.
These values should furthermore also be identical to those fitted from 
the spontaneous decays.
The identical functional form of the decay curves for spontaneous 
and photo-enhanced signals for an initially flat energy distribution is 
intuitively clear: Single photon absorption shifts up the energy distribution by the 
photon energy, but for a flat distribution this does not change its shape, except 
for an overall multiplicative constant related to the product of laser fluence and 
photon absorption cross section. 
As the decay profile is determined by the underlying energy distribution, a flat energy 
distribution also does not change the measured decay rate, apart from the multiplicative 
constant and the shift in the zero of time to $t_{\rm las}$.

\begin{figure}
\begin{center}
\includegraphics[width=0.7\columnwidth,angle=90]{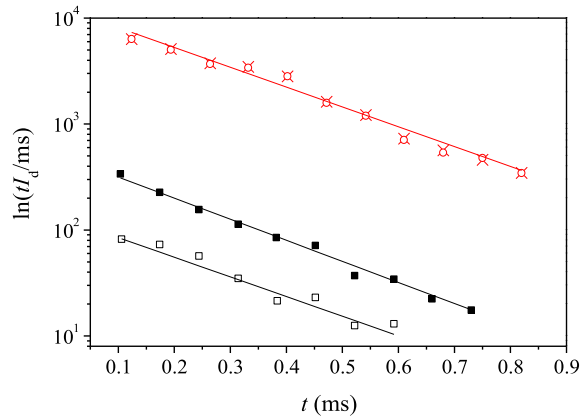}
\caption{\label{N19laser}
Plots of $\ln(tI_{\rm d})$ vs. time for C$_{19}^+$ for spontaneous decays of 
a spectrum without laser excitation (red open circles) and the almost
overlapping non-laser excited early time part of a laser-excited spectrum  (red crosses);
the photon enhanced decays for the laser firing times 10 ms (black 
filled squares) and for 80 ms (black open squares).
The time zero for these two curves are their respective laser firing times.
The point-by-point deviations from the straight line fits 
are mainly due to the betatron oscillations with an only very minor contribution 
from statistical fluctuations, demonstrating the reproducibility of the 
source.}
\end{center}
\end{figure}

Those requirements were tested previously in the study of the cationic carbon 
clusters of sizes reported in \cite{ChenPCCP2019}, but were also tested here 
with the laser excitation data recorded for C$_{11}^+$ and C$_{19}^+$.
Fig. \ref{N19laser} gives an example for C$_{19}^+$. 
The figure shows identical slopes for the spontaneous decay and the decays 
at the earliest and latest laser firing times, as required for the analysis.

\begin{figure}
\begin{center}
\vspace{-0.7cm}
\includegraphics[width=0.8\columnwidth,angle=90]{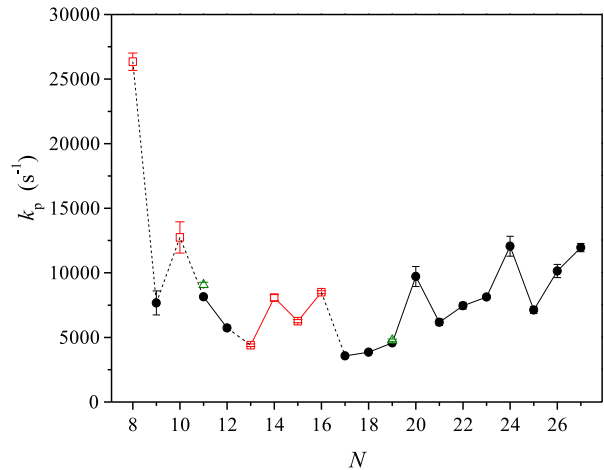}
\vspace{-0.7cm}
\caption{\label{kp-summary}
All measured values of $k_{\rm p}$,  shown as filled black circles for the 
spontaneous decays, and for $N=11$ and $N=19$ as olive triangles for the 
values from the photo-excitation experiments.
The clusters reported in \cite{ChenPCCP2019} are shown as open red squares.
Error bars are 1-$\sigma$ values of the averages of the independently 
analyzed measurements of each cluster size.}
\end{center}
\end{figure}

The fits for all storage times for both $N=19$ and $N=11$ are summarized 
in the averages shown in Fig. \ref{kp-summary} which shows the values for 
all measured cluster sizes. 
For reference also those reported in Ref. \cite{ChenPCCP2019} measured 
with the same storage ring and the same technique are given.

\section{Energies of emitting clusters}

The values of $k_{\rm p}$ provided by the analysis and shown in 
Fig. \ref{kp-summary} are those for which the radiative rate 
constant is equal to the unimolecular decay rate constant, 
\begin{equation}
\label{crossover}
k_{\rm p}(E) \approx k_{\rm p} = k_{\rm d}(E_{\rm crossing}),
\end{equation}
where $E_{\rm crossing}$ is the energy at the curve crossing point, and the 
approximation consists of setting $k_{\rm p}$ to an energy independent value.
The relation can be used to find the excitation energy of the clusters at this 
crossing point.
These energies are also the upper limit of the energies for which the photon 
emission is the dominant decay channel. 
To a good approximation the decays are all radiative below this energy and 
fragmentation above.

The value of $k_{\rm p}$ in Eq.\ref{crossover} is the measured value.
The dissociation rate constant, $k_{\rm d}(E)$, is computed with the parameters 
given by a density functional theory (DFT) calculation which provides 
ground state energies and vibrational frequencies as well as the 
geometric structures.
The calculations are performed with the ORCA 
4.2.1 software package (\cite{neese2018orca}).
The range-separated hybrid functional $\omega$B97X-D3 was employed 
(\cite{LinJCTC2013}), which also considers dispersion corrections 
(\cite{GrimmeJCP2010}).
For the calculations, all the electrons of carbon were considered 
implicitly, by the Def2-SVP basis set (\cite{WeigendPCCP2005}). 
This level of theory was selected in a previous study on C$_N^+$ 
clusters, where it was seen to correctly predict the transition 
from linear to cyclic conformers, at $N = 8$ (\cite{ChenPCCP2019}).
The calculations were conducted for $N = 9-27$, and for each 
of the cyclic and linear lowest-energy structures harmonic vibrational 
frequencies were calculated, which were always positive, 
confirming that the structures are minima and not saddle points.

The rate constant is calculated as the detailed balance value.
The ground states of the clusters in this study are cyclic or possibly 
polycyclic for the larger sizes (\cite{Helden1993}). 
We will proceed with the assumption that the monocyclic isomers 
represent all non-linear structures. 
The energies extracted from the decay dynamics derived in the following
should be fairly insensitive to the precise number of rings in the clusters. 
Furthermore, we will assume that the linear and cyclic isomers 
interconvert freely, i.e. that the fragmentation decay is slower than these 
two rates.

The transition state is taken as the detachment of a trimer 
(\cite{geusicJCP1987}) from a linear isomer. 
With no activation barrier for the reverse process of attachment and 
taking into account rotational degrees of freedom by summation 
over the trimer angular momentum states, the expression reads 
(\cite{book2ed})
\begin{eqnarray}\label{detbal}
k_{\rm d}(E)=\mkern-18mu
\underset{\text{0}}{\overset{E-D_{N,3}}\int}\mkern-18mu{\rm d} \epsilon 
~~\underset{\text{0}} {\overset{E-D_{N,3}-\epsilon}\int}\mkern-18mu{\rm d} 
\epsilon_{\rm rot}
~\omega(\epsilon)
\frac{\rho^{(\ell)}_{N-3,3}(E\!-\!D_{N,3}\!-\!\epsilon\! -\!\epsilon_{\rm rot})}
{\rho^{(c)}_N(E)} \!
\end{eqnarray}
with $D_{N,3}$ the dissociation energy and 
\begin{eqnarray}
\label{omega}
\omega(\epsilon) \equiv \rho_{3,\rm rot}(\epsilon_{\rm rot})
 \frac{\sigma_{N-3,3} \mu_{N-3,3}}{\pi^2 \hbar^3}  \epsilon. 
\end{eqnarray}
The numerically most important single parameter in this expression is the 
dissociation energy which is the difference between the ground state energies 
of the cyclic C$_N^+$ and the dissociation products C$_{N-3}^+ +$C$_3$.
It enters the energy argument of the level density of the decay product.
The vibrational level densities, $\rho^{(\ell)}_{N-3,3}(E)$ and $\rho^{(c)}_N(E)$
for the products (linear) and for the reactant (cyclic), 
respectively, were calculated with the Beyer-Swinehart algorithm (\cite{Beyer1973}).
$\rho^{(\ell)}_{N-3,3}(E)$ is calculated by pooling the vibrational frequencies
of C$_{N-3}^+$ and C$_3$. 
For the octamer, which is linear in the ground state and was measured in a 
previous experiment but also analyzed here, the equation is analogous, with 
the dissociation energy the linear-to-linear isomer value, and the cyclic isomer 
vibrational level density in the denominator replaced by that of the linear, 
$\rho^{(c)}(E) \rightarrow \rho^{(\ell)}(E)$. 

The reverse process capture cross section is set to 
$\sigma_{\rm geo} = 2 \pi r_1^2 \left(1 + 3^{1/3}\right)^2$, with $r_1=0.77$ \AA, 
i.e. half the typical carbon-carbon bond length.
The factor two represents the two possible attachment sites at each 
end of the linear cluster, and the remainder of the expression gives an estimate 
of the geometric size of the attachment cross section for each end of the chain.
$\mu_{N-3,3}$ is the reduced mass of the channel, and
$\rho_{3,{\rm rot}}$ is the neutral trimer rotational level density.
The two integrations were performed by direct summations of the discretized 
integrands over the kinetic energy released, $\epsilon$, and the rotational 
energy $\epsilon_{\rm rot}$.
For the trimer rotational states the quantum statistics of the 
rotational wave function was taken into account by using the symmetry 
number 2 and for the rotational constant the value $B=0.4306$ cm$^{-1}$ 
= 5.34 $\times 10^{-5}$ eV was used (\cite{SchmuttenmaerScience1990}).

\begin{figure}
\begin{center}
\vspace{-0.4cm}
\includegraphics[width=0.8\columnwidth,angle=90]{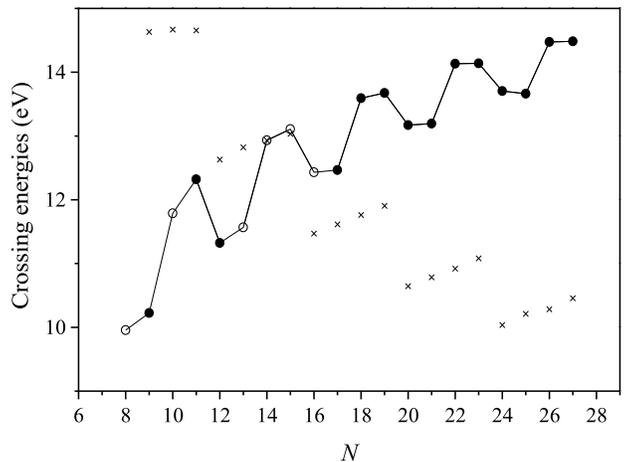}
\vspace{-0.7cm}
\caption{\label{E-limits} Cooling thresholds for the measured clusters 
(filled symbols) and previously measured clusters (open symbols).
The crosses give the second ionization energies determined with the 
DFT calculations described in the main text.
The experimental uncertainties are less than the symbols size.}
\end{center}
\end{figure}

Fig. \ref{E-limits} shows the limits calculated when the experimentally 
measured values of $k_{\rm p}$ are set equal to equation~(\ref{detbal}), which
is then solved for $E$.
The period of four is due to periodic filling of the $\pi$ electrons 
by addition of two electrons per carbon atom into four orbitals that 
are nearly degenerate.
A significant contribution to the energies in Fig. \ref{E-limits} is
the dissociation energies. 
As those also shape abundance spectra, the similarity of the curve seen here and
the quartet structure in the abundances seen in spectra produced in hot sources 
(\cite{JoyesJP1984,RohlfingJCP1984}) is not an accident, although the precise 
connection between abundances and stability is more complicated than a simple 
one-to-one relationship (\cite{book2ed}).
The general trend toward a higher energy for larger sizes is due in part to 
the kinetic shift, which in turn can be understood as due to the increase of 
the heat capacity with cluster size.

The figure also gives the second ionization energies calculated with the DFT 
software that was used for the other quantum chemical calculations in 
this work.
The values are similar to the values reported in \cite{Diaz-TenderoBJP2006}
for the smaller sizes where the two data sets overlap.
The lowest of the two values for crossing energy and second ionization 
energy in the figure sets a limit to the photon energies that after absorption will be 
re-emitted as strongly redshifted photons.
The cross-over energies in Fig. \ref{E-limits} exceed the second ionization 
energy of the largest species. 
This reduces the quantum efficiency for emission of lower energy photons 
for these species when photons with high energies are absorbed.

For a consideration of the complete picture it should be pointed out that absorption 
of above-threshold photons does not necessarily cause ionization.
Instead clusters and molecules may absorb the photons and undergo intramolecular 
vibrational relaxation, opening up for RF photon emission after energy dissipation.
Such non-ionizing photon absorption followed by relaxation has indeed been 
seen in C$_{60}$ (\cite{HansenPRL2017}).
The effect will cause the radiative cooling to be present also for photon energies 
above the second ionization threshold.
The precise mechanism behind the effect is still unknown, both qualitatively and 
concerning the precise branching ratios.
If active it will suppress a second ionization and raise the RF yield for carbon cation 
cluster sizes from $N=16$ and up. 

The derivation assumes that the photon emission rate constant is energy 
independent, and that the C$_3$ emission rate constant varies very rapidly at the 
point where the curves $k_{\rm d}$ and $k_{\rm ph}$ cross.
We will now consider both these approximations. 

A correction to the cross-over energy can be calculated in terms of the 
ratio of the photon energy ($h\nu$) and the evaporative activation energy 
($E_{\rm a}$) (\cite{ChenPCCP2019}). 
It changes the cross-over time and hence implicitly the cross-over energy with 
the factor
\begin{equation}
\left(1- \frac{h\nu}{E_{\rm a}}\right).
\end{equation}
This is small, and inverting the rate constant for this small time correction 
will result in an even smaller change in energy.

Secondly, the finite slopes of the two curves can potentially give rise to a smooth 
cross-over between the two channels as seen in \cite{JiJCP2017}.
The energy-specified photon emission branching ratio is given by the expression
\begin{equation}
R \equiv \frac{k_{\rm ph}}{k_{\rm d}+k_{\rm ph}}.
\end{equation}
The effect of finite values of the slopes of the two functions can be estimated by an 
expansion of the logarithm of the rate constant around the energy for which $R=1/2$. 
We have to first order for a generic activated process
\begin{eqnarray}
k(E + \delta E) \approx k(E) \exp\left( \delta E  \frac{\partial \ln k }{\partial E} \right).
\end{eqnarray}
The derivative can be estimated from the energy dependence of an activated process.
In analogy with the Van't Hoff equation, the logarithm can be written as
\begin{eqnarray}
\ln k = c - E_{\rm a}/T.
\end{eqnarray}
Taking the derivative with respect to energy gives 
\be
\frac{{\rm d}\ln k}{{\rm d} E} = \frac{E_{\rm a}}{CT^2},
\ee
where $C$ is the heat capacity ($k_{\rm B}$ is set to unity).
For the two parallel activated processes here, the relevant microcanonical 
temperatures are related approximately as (\cite{AndersenJCP2001})
\be
T_{\rm d} = T -\frac{E_{\rm d}}{2C}, ~~ T_{\rm ph} = T - \frac{h\nu}{2C}.
\ee
The channel-specific subtracted term is the leading order finite heat bath 
correction (\cite{Klots1989}).
As $h\nu < E_{\rm d}$, the ratio of the derivatives is therefore bounded as
\be
\frac{\frac{\partial \ln k_{\rm ph}}{\partial E}}
{\frac{\partial \ln k_{\rm d}}{\partial E}}
< \frac{h\nu}{E_{\rm d}}.
\ee
The values of $E_{\rm d}$ account for the breaking of two bonds and vary 
between 9.5 and 12.7 eV. 
Reasonably, the photon energies are significantly smaller than that, and then we 
can obtain a good estimate of the energy by setting the derivative of the photon 
emission rate constant equal to a constant in the relevant region.
Close to the expansion point $R=1/2$ we then have the ratio
\be
R \approx \frac{1}{1+\exp\left( \delta E  \frac{\partial \ln k_{\rm d}}{\partial E} \right)}.
\ee
The derivatives are very high for all clusters in the study, with values between 30/eV 
and 38/eV for $N=9-27$, and 110/eV for $N=1$, making the cross-over region some 
tens of meV wide.
For the lowest value of 30/eV, the branching ratio changes from 0.27 to 0.73 over 
the interval 0.066 eV.

The contrast to the very soft cross-over for anthracene found \cite{JiJCP2017}
is caused by the fact that for this molecule, both the derivative of the unimolecular 
reaction rate constant and the difference in the two derivatives are significantly smaller.

\section{Discussion}

The experiment did not record the decay channel, apart from the fact that the 
detected fragment is neutral.
It is still of some interest to know the moiety lost, however. 
Previous experiments with a range of different methods of production 
of the cations have established similar albeit not identical decay channels
to those used here (the loss of C$_3$).
The theory applied in the present work is in good agreement with the predominant 
channel seen in the photo-dissociation experiments in Ref. \cite{geusicJCP1987}.
The pattern reported in \cite{LifshitzIJMSIP1989} is as follows:
Loss of C$_1$ for C$_{11}^+$, C$_{12}^+$, whereas C$_{10}^+$ 
and C$_{16}^+$ all lose C$_3$, although for the latter this accounts for 
only half the yield and the remaining channels are emission of C$_N$, $N=1,2,5$.
Similar results were reported in \cite{RadiJCP1988}.
High energy collisions give similar results up to the measured largest cluster 
$N = 10$, as reported in \cite{BerofNIMB2009}.
A study using collision induced dissociation showed C$_5$ as the main loss 
channels for C$_{16}^+$ and C$_{19}^+$ (\cite{Sowa-ResatJPC1995}).
In the isomer specific study in (\cite{KoyasuCPL2012}) on C$_N^+$, $N=7-10$, 
the authors also find branching ratios where loss of the neutral trimer is 
dominant for both linear and cyclic isomers, most strongly for the linear ones, 
and with other channels present for both types.
The origin of the somewhat conflicting evidence from these previous 
experiments may partly be due to the different excitation procedures used.

In summary, the fragmentation losses appear to be predominantly of smaller 
molecules, larger than the monomer in the majority of cases and with the trimer 
as the most dominant. 
We have disregarded this scattered information on the channels in our analysis, 
primarily because they do not have a strong influence on the crossing energies, 
which will be determined mainly by the dissociation energies, with an additional 
kinetic shift (\cite{LifshitzEJMS2002}).
The exact nature of the fragment emission channel nevertheless deserves 
clarification, as the lowest energy channel is what ultimately competes with 
photon emission.
This clarification seems to be feasible with studies of delayed fragmentation in 
time-of-flight mass 
spectrometers, analogous to the procedure used in \cite{FerrariJCP2015} to 
resolve a very similar question for silicon clusters.

The data given here have small statistical uncertainties. 
The main uncertainty is the assumptions associated with the description of the 
fragmentation precursor, i.e. statistical mixing and the values for the dissociation 
energies.
Another uncertainty is the precise range of photon energies that will produce RF. 
This is of particular relevance in HI regions for $N \geq 16$ for which the second 
ionization potentials are below 13.6 eV, as non-ionizing absorption can give rise to 
thermally excited species even above 
threshold, and thus push the effective cross-over energy up. 
An interesting and connected side effect, which should be mentioned but will 
not be explored further here, is that absorption of high energy photons can also 
give rise to emission of low kinetic energy electrons, on the order of 1 eV 
(\cite{HansenPRL2017}).

Some other open questions remain for the clusters measured here.
The most important is the spectrum of the emitted photons.
The measured low temperature absorption spectra reported in 
Ref. (\cite{RademacherJPCA2022}) are likely to be smeared and possibly
shifted at the high excitation energies of the clusters in the present study.
This highly important question must be answered by future spectroscopic 
experiments.
Those will also be needed to answer the question of precise values of quantum 
efficiencies and of the time scales needed to reach energies where the dominant 
emission occurs via vibrational transitions.

\section{Conclusions}

This study has determined the radiative cooling rate constants of a range of 
cationic clusters to be above $10^{3}$ s$^{-1}$ and for some above 
$10^{4}$ s$^{-1}$.
These cluster-specific constants define the time scales at which fragmentation 
is quenched and cooling occurs predominantly through photon emission. 
The values for the crossing energies for the largest studied clusters are 
higher than the photon energies available in HI regions ($h\nu <13.6$ eV) so these 
largest clusters are expected to be stable in these regions, except when multi-photon
absorption occurs, cf. \cite{MontillaudAA2013}.
The energy at which this cross-over happens was calculated based on cluster 
properties given by a quantum chemical calculation and measured values for C$_3$.
This energy is significantly higher than any photon emitted.
The infrared emission rate constants that are much smaller than
the RF rates make that channel dominant only at correspondingly lower excitation 
energies. 
This leaves a fairly wide energy window where energy loss occurs mainly by 
emission of visible and near-infrared photons, very likely with a resulting
quantum efficiency above unity.
Although open questions remain on the decay channels, the results reported 
here therefore give quantitative evidence for a mechanism that can 
provide a high yield of photons in the visible and NIR spectral range.
Considering the difference in photon energies between the RF and IR 
photons and rates, the resulting RF energy dissipation rates for carbon cations at 
the crossing points are then higher than the infrared emission rate by three to four 
orders of magnitude.

The results pertain to the basic process of energy dissipation.
It thus provides input to calculations such as those reported in 
\cite{MontillaudAA2013}, where the long term time development of molecular 
abundances is analyzed.
The lifetime of clusters undergoing repeated cycles of photon absorption and emission
depend, in addition to the absorption cross section, on both the ambient 
vacuum ultraviolet (VUV) light intensity and the rate of energy dissipation.
The data here provide the first estimate of the difference in the fragmentation 
lifetimes of RF and IR dissipation.
For the most conservative estimate of a two-photon absorption event required to 
cause fragmentation, and using the values $10^{-4}$ s and $10^{-2}$ s for the 
RF and IR time scales gives a factor $10^4$ longer lifetimes.  
Clearly, the differences in radiative time scales of the two channels will have 
a corresponding impact on the limiting VUV flux for fragmentation.

Finally we note that the mechanism described here is present also
for larger molecules, for example fullerenes, both for neutral but in particular for 
cationic species with their relatively high second ionization potential 
(\cite{TomitaPRL2001}), as analyzed in \cite{BerneAA2015} in 
connection with the fullerene formation mechanism.
We also note that any excitation below the cross-over energy will give a quantum 
yield for RF photon emission of at least unity, because the first photon is emitted by 
that mechanism and the infrared cooling is far too weak to cause the rapid quenching.
The precise quantum yield remains a subject of further study, though.

\section*{Acknowledgements}

SI acknowledges the Grant-in-Aid for Early-Career Scientists (20K14386),
and PF acknowledges the support of the Research Foundation - Flanders 
(FWO)  with a postdoctoral grant. 
The computational resources and services used in this work were provided by the 
VSC (Flemish Supercomputer Center), funded by the Research Foundation - Flanders 
(FWO) and the Flemish Government.
KH acknowledges funding from National Science Foundation of China (NSFC) 
with the grant 12047501 and the 111 Project under 
Grant No. B20063 from the Ministry of Science and Technology of People's Republic of China.
Comments by A. Witt on astrophysical aspects and by T. Wakabayashi on carbon 
chemistry are gratefully acknowledged.
Several constructive comments by the reviewer are gratefully acknowledged.

\section*{Data Availability}
The data underlying this article will be shared on reasonable request to the 
corresponding author.


\end{document}